%% file: main.tex
\documentclass[sigconf, nonacm]{acmart} 

\AtBeginDocument{%
  }

\usepackage{amsmath}    
\usepackage{booktabs}   
\usepackage{multirow}   
\usepackage{graphicx}
\usepackage{array}
\usepackage{float}

\begin{document}

\title{When LLM-Inferred User Context Adds Value in Production Streaming Recommendation}

\author{Milad Sabouri}
\orcid{0009-0009-8219-5596}
\affiliation{%
   \institution{DePaul University / Comcast Technology AI}
   \city{Chicago}
   \state{IL}
   \country{USA}}
\email{msabouri@depaul.edu}
\email{milad_sabouri@comcast.com}

\author{Neeraj Sharma}
\orcid{0009-0008-7940-1523}
\affiliation{%
   \institution{Comcast Technology AI}
   \city{Sunnyvale}
   \state{CA}
   \country{USA}}
\email{neeraj_sharma@comcast.com}

\author{Sardar Hamidian}
\orcid{0009-0007-5377-0374}
\affiliation{%
   \institution{Comcast Technology AI}
   \city{Washington}
   \state{DC}
   \country{USA}}
\email{sardar_hamidian@comcast.com}

\author{Shaghayegh Agah}
\authornote{Corresponding author.}
\orcid{0000-0002-7063-9251}
\affiliation{%
   \institution{Comcast Technology AI}
   \city{Sunnyvale}
   \state{CA}
   \country{USA}}
\email{shaghayegh_agah@comcast.com}


\begin{abstract}
Contextual information in recommender systems is shifting from static, predefined variables toward latent representations inferred from behavior. Large language models support this shift by rendering an unstructured interaction history as a natural-language summary, which yields a thematic user context that can be encoded and used in place of an aggregate profile. The conditions under which such generated profiles outperform aggregate embeddings have received limited characterization mainly at the domain level. We evaluate semantic user-profiling strategies on a production streaming platform, ranking against the full catalog. The evaluation covers a 2×2 design space crossing representation type (aggregate or LLM-generated) with contextual scope (holistic history or attention-fused short-term and long-term contexts). The relative ordering of the two representation types is conditional on the user's consumption regime. Aggregate profiles are consistently stronger under habitual consumption, which characterizes approximately four-fifths of the population, while LLM-generated profiles are stronger for exploratory users whose subsequent interactions diverge semantically from their history. We also observe a popularity-attractor effect in LLM-generated profiles, which modestly raises within-list diversity while substantially lowering catalog coverage and reducing novelty. These results indicate that a context-aware system can select a profiling strategy from the inferred consumption regime rather than applying one representation to all users.
\end{abstract}



\maketitle

\input{body}

\bibliographystyle{ACM-Reference-Format}
\bibliography{ref}

\end{document}

%% file: body.tex
\section{Introduction}
\label{sec:introduction}

Context-aware recommender systems have traditionally represented context as a set of predefined variables, such as time of day, location, or device, that are specified in advance and observed directly. This representational view is limited by the fact that many of the factors shaping consumption are neither enumerable in advance nor directly observable. An alternative is to infer context from the behavioral record itself, treating it as a latent quantity to be estimated rather than measured.

Large language models make this alternative practical for content-based recommendation. Given a user's interaction history and the text describing the items in it, an LLM can produce a natural-language summary of that user's preferences. Encoding the summary with a sentence encoder yields a vector that the downstream architecture consumes like any other user profile, so an inferred context can substitute directly for the conventional representation, which aggregates item embeddings into a centroid~\cite{lops2011content}. Either representation can be built over the full history or over a partition of it into recent and historical segments fused by attention~\cite{bahdanau2014neural, vaswani2017attention}. These two choices, the type of representation and the contextual scope over which it is constructed, define the design space in Table~\ref{tab:design_grid}.

This space has been examined on public e-commerce benchmarks. An LLM-generated profile of recent and historical segments, fused by attention, outperformed both a centroid profile and an aggregate profile of the same segmented form, with the margin varying across
domains~\cite{sabouri2025using}. Gains were pronounced on a movie and television catalog and receded on a games catalog, where the aggregate segmented profile led on three of the four metric and cutoff combinations. That line of work read the difference as a property of the domain and proposed applying LLM profiling selectively as a direction
rather than testing it~\cite{sabouri2025effectiveness}. Its evaluations ranked against catalogs of a few thousand to roughly fourteen thousand items, measured accuracy alone, and located conditionality between datasets rather than among the users inside one.

\begin{table}[t]
  \caption{The 2$\times$2 design space compared in this work. Rows are representation type, columns contextual scope.}
  \label{tab:design_grid}
  \small
  \begin{tabular}{@{}p{0.16\linewidth} p{0.37\linewidth} p{0.37\linewidth}@{}}
    \toprule
    & \textbf{Holistic} (full history) & \textbf{Disentangled} (short/long split, attention-fused) \\
    \midrule
    \textbf{Aggregate} & \textbf{Centroid (C)}: SBERT item-embedding centroid over the full history. & \textbf{TemporalCentroid (TC)}: short- and long-term centroids, attention-fused. \\
    \addlinespace[2pt]
    \textbf{LLM-generated} & \textbf{Narrative (N)}: one LLM summary of the full history, SBERT-encoded. & \textbf{TemporalNarrative (TN)}: two LLM summaries (recent, full), SBERT-encoded, attention-fused. \\
    \bottomrule
  \end{tabular}
\end{table}

We ask whether that result carries to a production setting, and whether the conditionality it identified also operates at a finer grain. The empirical results analyzed here are drawn from the same production dataset and experimental framework as our broader study of LLM-based user profiling~\cite{sabouri2026llm}, but here we examine them through a context-aware recommendation lens, focusing on consumption regimes and their implications for context-dependent profile selection. We compare the same four profiling strategies, crossing representation type with contextual scope, on a sample from a production streaming platform whose filtered catalog is more than an order of magnitude larger, ranking against that catalog in full rather than against a sampled candidate pool. We hold the item encoder, the scoring head, and the training protocol fixed so that the profile is the only variable, report beyond-accuracy metrics alongside accuracy, and analyze users separately according to whether their subsequent consumption revisits the content already present in their history.

Three findings follow. First, whether aggregate or LLM-generated context produces better recommendations depends on the user's consumption regime. Aggregate profiles are stronger for habitual users, who form the large majority of the population, while LLM-generated context is stronger for exploratory users, whose subsequent interactions diverge semantically from their history. Second, LLM-generated profiles act as a popularity attractor. Individual recommendation lists become modestly more diverse, while the catalog footprint across users contracts and recommended items become more popular. Third, the effect of contextual scope depends on the representation it is applied to. It changes little for aggregate profiles. For LLM-generated profiles, it reduces the shortfall against the aggregate baseline on habitual users without removing it, and reduces the advantage on exploratory users at the top of the list. Together these results indicate that a context-aware system can select its profiling strategy from the inferred consumption regime rather than applying one representation to every user, and we report a preliminary analysis of how far that regime is predictable before the consumption it describes is observed.


\section{Related Work}
\label{sec:related_work}

\paragraph{Context in recommender systems}
\label{sec:related_context}

Context-aware recommendation~\cite{Adomavicius2022} has largely treated context as a set of variables fixed in advance and observed directly, such as time, location, companion, or device~\cite{kaminskas2012contextual, otebolaku2015context}, an approach that a systematic review of the area finds to dominate the literature~\cite{villegas2018characterizing}. Because the variables are chosen by the designer, the factors a system can respond to are bounded by what was anticipated. Recent work uses language models to widen that bound. Geo-temporal context enrichment prompts an LLM with a timestamp and a coarse location and returns embeddings that carry holidays, seasons, and local events without a curated calendar, and reports that the resulting gains vary enough across settings to motivate adaptive strategies~\cite{kim2025time}. The source variables there remain the familiar ones and the language model supplies the semantics attached to them. We infer the context from the behavioral record instead, where no variable is designated in advance.

\paragraph{Aggregate and LLM-generated user profiles}
\label{sec:related_profiling}

Content-based recommendation has long built user representations by aggregating item-level features~\cite{lops2011content}. With sentence encoders such as SBERT~\cite{reimers-2019-sentence-bert}, items are placed in a continuous space and profiles are formed as centroids or other pooling functions over item embeddings, with weighted aggregation, recency-decayed pooling, and attention-based fusion explored as refinements. A more recent line replaces pooling with an LLM that renders the history as a natural-language summary~\cite{lubos2024llm, ma2024xrec}, which a sentence encoder converts into a vector for the same downstream architecture. Such profiles have been applied to explainable and sequential recommendation and to user understanding, and evaluated mainly on academic benchmarks~\cite{sabouri2025towards}.

\paragraph{Temporal structure in user modeling}
\label{sec:related_temporal}

Sequence models, including recurrent~\cite{tan2016improved}, self-attentive~\cite{kang2018self}, and bidirectional-transformer~\cite{sun2019bert4rec} recommenders, treat the history as an ordered sequence and predict the next item conditional on it. A complementary line decomposes behavior into short-term and long-term components~\cite{zhu2017what, wang2019modeling} and fuses them by attention, with the weighting learned from data~\cite{bahdanau2014neural, vaswani2017attention}. Recent work applies that decomposition to LLM-generated profiles~\cite{sabouri2025using}, and we adopt its prompting and fusion patterns for the disentangled narrative strategy in Section~\ref{sec:temporal_narrative}. Both disentangled strategies compared here follow this framework within content-based ranking.


\section{Methodology}
\label{sec:methodology}

We compare four semantic user-profiling strategies for content-based recommendation, crossing two design axes. \emph{Representation type} separates profiles derived as numerical aggregates of item embeddings from profiles generated as natural-language summaries by an LLM. \emph{Contextual scope} separates profiles built over the full chronological history from profiles that partition it into recent and historical segments fused by attention. The four strategies, \emph{Centroid} (C), \emph{TemporalCentroid} (TC), \emph{Narrative} (N), and \emph{TemporalNarrative} (TN), are organized in Table~\ref{tab:design_grid}.

All four follow definitions established on public benchmarks~\cite{sabouri2025using}. Centroid corresponds to the content-based baseline in that work, TemporalCentroid to its temporal-fusion baseline, TemporalNarrative to the LLM temporal profiling method it introduced, and Narrative to the variant of that method with the temporal split removed, reported there as an ablation. We evaluate all four as strategies in their own right. They share a downstream architecture and differ only in the construction of the profile vector $\mathbf{p}_u$, which isolates profile design as the sole experimental variable.

\subsection{Notation and scoring}
\label{sec:notation}

For a user $u$, let $\mathcal{H}_u = \{(i, t)\}$ denote the chronologically ordered interaction history over catalog $\mathcal{I}$, where each item $i$ carries textual metadata $m_i$ comprising its title and description. The task is to score every catalog item and return the top-$K$:
\begin{equation}
  s(u, i) = f_\theta\!\left(\mathbf{p}_u \,\Vert\, \mathbf{e}_i\right),
  \label{eq:scoring}
\end{equation}
where $\Vert$ is concatenation, $f_\theta$ a learned multi-layer perceptron (MLP), $\mathbf{p}_u \in \mathbb{R}^d$ the profile vector produced by one of the four strategies below, and $\mathbf{e}_i = \text{SBERT}(m_i)$ the item embedding~\cite{reimers-2019-sentence-bert}.

\subsection{Profiling strategies}
\label{sec:strategies}

\subsubsection{Centroid (C)}
\label{sec:centroid}

The Centroid profile is the arithmetic mean of SBERT embeddings of all items in the history,
\begin{equation}
\mathbf{p}^{C}_u = \frac{1}{|\mathcal{H}_u|} \sum_{(i, t) \in \mathcal{H}_u} \text{SBERT}(m_i),
\end{equation}
following conventional centroid-based content profiling~\cite{lops2011content}. It uses no temporal structure and produces no natural-language artifact, and serves as the reference against which the other strategies are compared.

\subsubsection{TemporalCentroid (TC)}
\label{sec:temporal_centroid}

TemporalCentroid computes two centroids over different temporal segments, following the temporal-fusion baseline of~\cite{sabouri2025using}. Given a short-term window $\rho \in (0, 1)$, the short-term segment $\mathcal{H}^{\text{short}}_u$ holds the most recent $\rho \cdot |\mathcal{H}_u|$ interactions. The long-term segment is the full history $\mathcal{H}_u$ rather than the complement of the short-term segment, a choice carried over from that work, where the long-term profile is built over the complete interaction record. Retaining recent interactions in both segments allows the long-term profile to capture persistent themes anchored by all of the user's behavior. The segment centroids $\mathbf{c}^{\text{short}}_u$ and $\mathbf{c}^{\text{long}}_u$ are combined by attention~\cite{wang2019modeling, bahdanau2014neural}:
\begin{equation}
\label{eq:tc}
\mathbf{p}^{TC}_u = \alpha_u \cdot \mathbf{c}^{\text{short}}_u + (1 - \alpha_u) \cdot \mathbf{c}^{\text{long}}_u,
\end{equation}
where $\alpha_u \in (0, 1)$ is a softmax-normalized scalar produced from the two centroids by a learnable single-layer attention parameterized by $\mathbf{W}_a \in \mathbb{R}^{1 \times d}$, as formulated in~\cite{sabouri2025using}.

\subsubsection{Narrative (N)}
\label{sec:narrative}

Narrative removes the temporal split, prompting the LLM once over the full history for a single description $T^{N}_u$ of the user's preferences. A simplified prompt template is ``\textit{Given the user's complete viewing history below, produce a concise summary of the user's overall preferences, emphasizing enduring themes and recurring patterns across the full history.}'' The summary is encoded by the same SBERT model used for items,
\begin{equation}
\mathbf{p}^{N}_u = \text{SBERT}(T^{N}_u).
\end{equation}
This corresponds to the ablation of~\cite{sabouri2025using} in which all interactions collapse into a single textual summary.

\subsubsection{TemporalNarrative (TN)}
\label{sec:temporal_narrative}

TemporalNarrative is the LLM temporal profiling method introduced in~\cite{sabouri2025using}, which combines LLM-generated representation with temporal scope. Using the same window $\rho$ and segment definitions as TemporalCentroid, the LLM is prompted twice, once over the short-term segment and once over the full history. We adapt the prompt wording of that work to the streaming domain, asking for recent shifts in genre, tone, or content type in the first case and for themes that remain stable across the history in the second. The summaries $T^{\text{short}}_u$ and $T^{\text{long}}_u$ are SBERT-encoded and fused by the attention of Equation~\eqref{eq:tc}:
\begin{equation}
\mathbf{p}^{TN}_u = \alpha_u \cdot \text{SBERT}(T^{\text{short}}_u) + (1 - \alpha_u) \cdot \text{SBERT}(T^{\text{long}}_u).
\end{equation}

\subsection{Shared components}
\label{sec:shared}

All other components are held constant. Item embeddings come from a frozen pretrained SBERT model applied to each item's title and description, precomputed and cached. The scoring head is a two-layer ReLU MLP over $\mathbf{p}_u \,\Vert\, \mathbf{e}_i$, as in Equation~\eqref{eq:scoring}. Each strategy is trained with binary cross-entropy under uniform negative sampling, with five negatives per positive, and optimized with Adam~\cite{kingma2014adam}. 


\begin{table}[t]
  \caption{Accuracy results at $\text{PCT} = 25\%$. All values are percentage change relative to Centroid, which is the within-segment, within-$K$ reference and is therefore omitted from the rows. $^{\ast}$ indicates the difference from Centroid is statistically significant at $\alpha = 0.05$.}
  \label{tab:headline}
  \scriptsize
  \setlength{\tabcolsep}{4pt}
  \begin{tabular*}{\linewidth}{@{\extracolsep{\fill}}l l rrr rrr@{}}
    \toprule
    & & \multicolumn{3}{c}{$K = 10$} & \multicolumn{3}{c}{$K = 100$} \\
    \cmidrule(lr){3-5} \cmidrule(lr){6-8}
    Segment & Strategy & Recall & NDCG & HR & Recall & NDCG & HR \\
    \midrule
    \multirow{3}{*}{All}
    & TemporalCentroid  & $+0.5$  & $+4.7^{\ast}$  & $+1.4$  & $-0.3$ & $+2.9^{\ast}$  & $+0.3$ \\
    & Narrative         & $-30.7^{\ast}$ & $-26.9^{\ast}$ & $-20.9^{\ast}$ & $-9.0^{\ast}$ & $-11.4^{\ast}$ & $-6.1^{\ast}$ \\
    & TemporalNarrative & $-21.6^{\ast}$ & $-15.2^{\ast}$ & $-14.2^{\ast}$ & $-9.0^{\ast}$ & $-11.5^{\ast}$ & $-5.0^{\ast}$ \\
    \midrule
    \multirow{3}{*}{Non-Explorer}
    & TemporalCentroid  & $+0.7$  & $+4.8^{\ast}$  & $+1.7^{\ast}$  & $-0.3$  & $+3.0^{\ast}$  & $+0.3$ \\
    & Narrative         & $-31.7^{\ast}$ & $-27.8^{\ast}$ & $-21.8^{\ast}$ & $-10.2^{\ast}$ & $-11.9^{\ast}$ & $-7.2^{\ast}$ \\
    & TemporalNarrative & $-22.4^{\ast}$ & $-15.8^{\ast}$ & $-14.7^{\ast}$ & $-10.2^{\ast}$ & $-12.3^{\ast}$ & $-5.9^{\ast}$ \\
    \midrule
    \multirow{3}{*}{Explorer}
    & TemporalCentroid  & $-7.4$  & $-6.0$  & $-10.2$ & $-0.7$ & $-0.2$  & $+0.5$ \\
    & Narrative         & $+18.7$ & $+31.5^{\ast}$ & $+14.8$ & $+9.5^{\ast}$ & $+12.8^{\ast}$ & $+6.2$ \\
    & TemporalNarrative & $+14.3$ & $+21.8$ & $+9.0$  & $+11.4^{\ast}$ & $+10.2^{\ast}$ & $+7.1$ \\
    \bottomrule
  \end{tabular*}
\end{table}

\section{Experiments}
\label{sec:experiments}

\subsection{Experimental setup}
\label{sec:setup}

\subsubsection{Dataset and sample}
\label{sec:dataset}

The dataset is sampled from a production streaming platform serving Movies, TV Shows, and Sports content. We evaluate a random sample of 10{,}000 users. The catalog is filtered to items whose title and description are in English and carry enough text for meaningful encoding, leaving approximately $2.4 \times 10^{5}$ items. Each user's interaction history is split chronologically into 80\% training, 10\% validation, and 10\% test, per user,
which prevents within-user temporal leakage.

\subsubsection{Scope}
\label{sec:scope}

All four strategies are content-based. We exclude sequence-based and collaborative methods such as SASRec~\cite{kang2018self} and BERT4Rec~\cite{sun2019bert4rec}, which address the complementary problem of next-item prediction from collaborative signal, and comparing them
here would confound signal source with profile design. Holding the paradigm fixed lets the 2$\times$2 contrast isolate profile design.

\subsubsection{Consumption regimes}
\label{sec:segmentation}

Profiling effectiveness may depend on whether a user's subsequent consumption resembles past behavior or departs from it. We label a user an \textit{Explorer} if no item in the test history appears in the training history, and a \textit{Non-Explorer} otherwise, and refer to the two as consumption regimes. Non-Explorers reinforce established preferences, while Explorers occupy a regime in which past consumption has limited predictive power for what follows. The sample contains 1{,}761 Explorers, approximately 18\% of users. The label is post-hoc and analytical, and Section~\ref{sec:routing_implications} considers how an inference-time proxy might be built. Results are reported for the full population (All) and
for each regime.

\subsubsection{Evaluation metrics}
\label{sec:metrics}

We report six metrics at $K \in \{10, 100\}$. Recall@$K$ is the fraction of test items in the top-$K$ list, NDCG@$K$ its rank-aware variant, and HitRate@$K$ the fraction of users with at least one test item in the top-$K$. Diversity@$K$ is the average pairwise cosine dissimilarity within a user's top-$K$ list~\cite{castells2015novelty}. Coverage@$K$ is the fraction of the catalog appearing in any user's top-$K$ list. Novelty@$K$ is the mean self-information of recommended items, $\frac{1}{K}\sum_{i \in R_u^K} -\log_2 p(i)$, where $R_u^K$ is the user's top-$K$ list and $p(i)$ is the fraction of training users with at least one interaction with item $i$~\cite{vargas2011rank}. Lower Novelty indicates concentration on more popular items.

\subsubsection{Short-term window}
\label{sec:pct_sweep}

The window $\rho$ governs the partition in TemporalCentroid and TemporalNarrative (Section~\ref{sec:temporal_centroid}). We refer to it as PCT, the percentage of the history assigned to the short-term segment, and sweep $\rho \in \{0.15, 0.20, 0.25, 0.30\}$, reported as percentages. Centroid and Narrative are PCT-invariant by construction.

\subsubsection{Reporting and statistical testing}
\label{sec:reporting}

Platform confidentiality requires that values be reported as relative percentage change against a within-segment, within-$K$ reference. For cross-model comparisons that reference is Centroid at the same segment and cutoff. For the sensitivity analysis in Section~\ref{sec:results_pct}, each model is normalized to its own value at $\rho = 0.25$ in the same segment, which isolates the shape of its response to PCT from level differences across models. We test pairwise per-user metric values with the Wilcoxon signed-rank test~\cite{wilcoxon_1945_individual, hollander_2014_nonparametric}, independently for each metric, cutoff, and regime at PCT~=~25\%. Significant differences at $\alpha = 0.05$ are marked with an asterisk in Tables~\ref{tab:headline} and~\ref{tab:beyond_accuracy}. Coverage is computed across all users and is therefore not subject to per-user testing.

\subsection{Results}
\label{sec:results}

\subsubsection{Accuracy by regime}
\label{sec:results_accuracy}

Table~\ref{tab:headline} reports accuracy at PCT~=~25\%, a value inside the stable range identified in Section~\ref{sec:results_pct}. The ordering of the two representation types depends on the regime. On \textit{All}, both LLM-generated strategies fall below Centroid on Recall@10, by 30.7\% for Narrative and 21.6\% for TemporalNarrative, while TemporalCentroid is within half a point of it. The pattern holds at $K = 100$ and across NDCG and HitRate, and every LLM comparison is significant. Because Non-Explorers are approximately 82\% of the sample, \textit{All} tracks them closely. On \textit{Explorer} the ordering inverts. Narrative leads Centroid by 18.7\% on Recall@10 and TemporalNarrative by 14.3\%, while TemporalCentroid falls 7.4\% below. The direction holds across NDCG and HitRate and at $K = 100$, though significance is only partial. Recall and NDCG are significant for both LLM strategies at $K = 100$, and NDCG at $K = 10$ for Narrative, while HitRate reaches significance at neither cutoff. We attribute this to the smaller Explorer sample. 

The scope axis behaves differently under each representation. TemporalCentroid stays within a few points of Centroid throughout. Under the LLM representation the split matters, and its sign depends on the regime. On \textit{All} and \textit{Non-Explorer}, TemporalNarrative leads Narrative by approximately 9 points of Recall@10 and 11 to 12 points of NDCG@10, without approaching Centroid. On \textit{Explorer} at $K = 10$ the order reverses and Narrative leads TemporalNarrative on all three metrics. At $K = 100$ the two fall within about three points and their order is not stable.

\begin{table}[t]
  \caption{Beyond-accuracy results at $\text{PCT} = 25\%$. All values are percentage change relative to Centroid, which is the within-segment, within-$K$ reference and is therefore omitted from the rows. $^{\ast}$ indicates significance at $\alpha = 0.05$. $^{\dagger}$ Coverage is a system-level metric and is not subject to per-user statistical testing.}
  \label{tab:beyond_accuracy}
  \scriptsize
  \setlength{\tabcolsep}{4pt}
  \begin{tabular*}{\linewidth}{@{\extracolsep{\fill}}l l rrr rrr@{}}
    \toprule
    & & \multicolumn{3}{c}{$K = 10$} & \multicolumn{3}{c}{$K = 100$} \\
    \cmidrule(lr){3-5} \cmidrule(lr){6-8}
    Segment & Strategy & Div. & Cov.$^{\dagger}$ & Nov. & Div. & Cov.$^{\dagger}$ & Nov. \\
    \midrule
    \multirow{3}{*}{All}
    & TemporalCentroid  & $+0.0$ & $-5.8$  & $+1.9^{\ast}$  & $-0.1^{\ast}$ & $-8.7$  & $+0.5^{\ast}$  \\
    & Narrative         & $+4.2^{\ast}$ & $-79.8$ & $-17.8^{\ast}$ & $+3.0^{\ast}$ & $-82.2$ & $-19.0^{\ast}$ \\
    & TemporalNarrative & $+4.0^{\ast}$ & $-79.2$ & $-19.9^{\ast}$ & $+3.2^{\ast}$ & $-83.7$ & $-20.5^{\ast}$ \\
    \midrule
    \multirow{3}{*}{Non-Explorer}
    & TemporalCentroid  & $+0.1$ & $-5.9$  & $+1.6^{\ast}$  & $-0.0$ & $-8.7$  & $+0.3^{\ast}$  \\
    & Narrative         & $+4.1^{\ast}$ & $-78.3$ & $-17.5^{\ast}$ & $+3.0^{\ast}$ & $-81.1$ & $-18.7^{\ast}$ \\
    & TemporalNarrative & $+4.0^{\ast}$ & $-77.6$ & $-19.6^{\ast}$ & $+3.2^{\ast}$ & $-82.7$ & $-20.3^{\ast}$ \\
    \midrule
    \multirow{3}{*}{Explorer}
    & TemporalCentroid  & $-0.1$ & $-4.1$  & $+2.9^{\ast}$  & $-0.2^{\ast}$ & $-5.9$  & $+1.1^{\ast}$  \\
    & Narrative         & $+4.3^{\ast}$ & $-74.0$ & $-18.9^{\ast}$ & $+3.1^{\ast}$ & $-77.2$ & $-20.7^{\ast}$ \\
    & TemporalNarrative & $+3.9^{\ast}$ & $-74.0$ & $-21.1^{\ast}$ & $+3.1^{\ast}$ & $-79.2$ & $-21.6^{\ast}$ \\
    \bottomrule
  \end{tabular*}
\end{table}

\subsubsection{Beyond-accuracy}
\label{sec:results_beyond}

Table~\ref{tab:beyond_accuracy} reports Diversity, Coverage, and Novelty. The LLM-generated strategies raise within-list Diversity over Centroid by approximately 4\% at $K = 10$ and 3\% at
$K = 100$, lower Coverage by approximately 80\%, so recommendations across users concentrate on a smaller part of the catalog, and lower Novelty by 18 to 22\%, so recommended items are more popular. TemporalCentroid stays within a few percent of Centroid on all three. The direction is the same across different PCT settings and all three regimes, and the Coverage and Novelty gaps widen at $K = 100$. Every per-user Novelty and Diversity comparison between an LLM-generated and an aggregate strategy is significant at $\alpha = 0.05$ across both cutoffs and all three regimes, which makes this the most consistent effect in our results.

\subsubsection{Sensitivity to the short-term window}
\label{sec:results_pct}


We sweep PCT and measure Recall@10 for \textit{All} and \textit{Explorer}, with each temporal strategy expressed relative to its own value at PCT~=~25\%. Centroid and Narrative are PCT-invariant by construction and hold the zero reference, so the sweep concerns only TemporalCentroid and TemporalNarrative. On \textit{All}, both vary by under 4\% across PCT $\in \{15\%, 20\%, 25\%\}$, and at 30\% TemporalNarrative falls to approximately $-4.7\%$ while TemporalCentroid holds near its reference. The \textit{Explorer} regime is more sensitive. TemporalNarrative falls to approximately $-9.5\%$ at 15\%, recovers between 20\% and 25\%, and falls to approximately $-19\%$ at 30\%, while TemporalCentroid is flat until dropping to approximately $-11\%$ at 30\%. That both temporal strategies degrade most at 30\% on \textit{Explorer} suggests the window grows large enough that the partition stops separating recent from historical behavior for the users whose recent behavior departs from their history. We adopt PCT~=~25\% for the reported results. 


\section{Discussion}
\label{sec:discussion}

\subsection{Interpreting the regime-conditional effect}
\label{sec:mechanism}

A Centroid profile is the mean of the embeddings of consumed items, so it sits near those items in embedding space. When subsequent consumption draws from the same neighborhood, that specificity is an advantage, since ranking against item embeddings recovers similar items. This is the habitual regime, where aggregate profiles are stronger. An LLM-generated profile is positioned differently. The model compresses the history into a summary that abstracts from individual items toward themes, and the encoding of that summary is placed according to the language of the summary rather than the items that informed it. For exploratory users, item-level affinity predicts little, and a representation general enough to reach unseen items is more useful. Thematic abstraction supplies that, which is one explanation for the ordering in this regime.

The popularity-attractor pattern admits a related explanation. The LLM is trained on text in which popular items are discussed more often than long-tail items, giving denser semantic associations for popular content, and the sentence encoder is trained on similar distributions. Summaries encoded into this space tend toward regions occupied by the embeddings of popular items. This explanation predicts the observed combination, since thematic abstraction raises within-list diversity while popularity-skewed regions lower novelty and coverage, and it is consistent with popularity bias reported for language models~\cite{navigli_2023_biases, gallegos_2024_bias_survey} and for embedding-based retrieval~\cite{abdollahpouri_2019_popularity_bias}. We do not separate the contributions of generation and encoding, so this remains a hypothesis, although the consistency of the pattern across settings and segments suggests it is not an artifact of one configuration.

The scope axis behaves differently under the two representations. For aggregate profiles it changes little, since TemporalCentroid stays close to Centroid on the full sample. For LLM-generated profiles it matters. On habitual users, TemporalNarrative recovers roughly nine points of Recall@10 over Narrative while remaining well short of the aggregate baseline, and on exploratory users at $K = 10$ the ordering reverses and Narrative leads. One reading is that a summary restricted to recent items is narrower and closer to the items themselves than a summary of the whole history, so the split makes an LLM-generated profile more concrete. Concreteness is what the habitual regime rewards and what the exploratory regime does not, which places this result on the same footing as the representation result above. The direction on the full sample agrees with the earlier ablation, where removing the temporal split degraded accuracy on a public benchmark~\cite{sabouri2025using}. What the regime-level view adds is that the direction holds only for the majority regime.

\subsection{Context-adaptive strategy selection}
\label{sec:routing_implications}

If the user's regime, exploratory or habitual, can be estimated at serving time, a system can select the representation matched to it rather than applying one representation to every user, which treats the profiling strategy as itself context-dependent, a direction proposed on the basis of domain-level differences in earlier work~\cite{sabouri2025effectiveness}. The regime that benefits from LLM-generated context covers approximately 18\% of users in our sample, so the choice is not a marginal one.

The Explorer label is post-hoc, computed from the relationship between training and test items, so such a system depends on a separate prediction problem. To assess whether that prediction is plausible, we tested whether training-period features correlate with the label using point-biserial correlation~\cite{tate_1954_pointbiserial, hollander_2014_nonparametric}. Two are significant, history size ($r = -0.30$, $p < 0.001$) and within-history content diversity, measured as the mean pairwise cosine distance across training items ($r = +0.11$, $p < 0.001$). Users with smaller and more diverse histories are more often Explorers. The regime is therefore partially inferable from information available before the consumption it describes, although this is an observation about feasibility rather than a predictor.

Two consequences follow. When the estimate is unavailable or uncertain, defaulting to the aggregate representation is the conservative choice, since it is stronger on the majority, while selection concentrates the popularity-attractor effect on exploratory users, lowering coverage and novelty for the population a system might otherwise use to broaden exposure. Contextual adaptation of this kind therefore redistributes a bias rather than removing it.

\subsection{Limitations}
\label{sec:limitations}

The evaluation covers one platform, one content domain, and one generation and encoding pipeline. Since the finding reported here is itself a conditional one, other settings may condition differently, and the regimes identified in this sample should not be assumed to transfer unchanged. The earlier evaluations on public data did not separate the two regimes, so whether the domain-level differences reported there reduce to differences in regime composition remains untested~\cite{sabouri2025using, sabouri2025effectiveness}. The evaluation is offline, while the selection mechanism it motivates would operate online.


\section{Conclusion}
\label{sec:conclusion}

We compared four semantic user-profiling strategies, crossing representation type with contextual scope, on a sample from a production streaming platform. The relative effectiveness of aggregate and LLM-generated context was conditional on the user's consumption regime. Aggregate profiles were stronger for the habitual majority, while LLM-generated context was stronger for exploratory users whose subsequent consumption diverges from their history. Contextual scope changed little under the aggregate representation and mattered under the LLM one, where its sign also depended on the regime. Beyond accuracy, LLM-generated profiles concentrated recommendations on a smaller and more popular part of the catalog. Training-period features correlate with the regime, which suggests it is partially inferable before the consumption it describes is observed. These results support treating the profiling strategy as context-dependent, with a system inferring the regime a user is in and selecting the representation matched to it rather than applying one representation to every user.